*Fatemeh Alizadeh, Dave Randall, Peter Tolmie, Minha Lee, Yuhui Xu, Sarah Mennicken, Mikołaj P. Woźniak, Dennis Paul, and Dominik Pins (2024): ECSCW 2024 Future of Home-living: designing Smart Home Spaces for Modern domestic Life. In: Proceedings of the 22nd European Conference on Computer-Supported Cooperative Work: The International Venue on Practice-centered Computing on the Design of Cooperation Technologies – Workshop Proposal, Reports of the European Society for Socially Embedded Technologies (ISSN 2510-2591), DOI: 10.48340/ecscw2024_ws02*

# Future of Home-living: Designing Smart Spaces for Modern Domestic Life

Fatemeh Alizadeh, Peter Tolmie
University of Siegen
Fatemeh.alizadeh@uni-siegen.de, Peter.tolmie@uni-siegen.de

Dave Randall
University of Siegen
Daverandall2008@gamail.com

Minha Lee, Yuhui Xu
Eindhoven University of Technology
m.lee@tue.nl, y.xu1@tue.nl

Sarah Mennicken
DREI Solutions
sarah@drei-solutions.com

Mikołaj P. Woźniak
University of Oldenburg
mikolaj.wozniak@uni-oldenburg.de

Dennis Paul, Dominik Pins
Fraunhofer Institute
Dennis.paul@fit.fraunhofer.de, dominik.pins@fit.fraunhofer.de**Abstract.** The evolution of smart home technologies, particularly agentic ones such as conversational agents, robots, and virtual avatars, is reshaping our understanding of home and domestic life. This shift highlights the complexities of modern domestic life, with the household landscape now featuring diverse cohabiting units like co-housing and communal living arrangements. These agentic technologies present specific design challenges and opportunities as they become integrated into everyday routines and activities. Our workshop envisions smart homes as dynamic, user-shaped spaces, focusing on the integration of these technologies into daily life. We aim to explore how these technologies transform household dynamics, especially through boundary fluidity, by uniting researchers and practitioners from fields such as design, sociology, and ethnography. Together, we will develop a taxonomy of challenges and opportunities, providing a structured perspective on the integration of agentic technologies and their impact on contemporary living arrangements.2

# Introduction and Motivation

The questions regarding the meaning of "home" and people' exaptation's of their ideal future home have been around since 2003 (Eggen et al., 2003). In recent years, there has been a considerable and burgeoning interest in 'smart home' technologies. Nevertheless, the technology is developing as fast, if not faster, than academic output. Today, various forms of agentic technologies, whether ChatGPT, robots, or virtual avatars, are making their way into our homes, playing various roles in everyday home life (e.g., (Heiyanthuduwa et al., 2020; Koomsap et al., 2023; Seymour, 2020; Urquhart et al., 2019)). This makes the exploration of their impact and integration into domestic settings more pertinent than ever.

Historically, homes have been recognized as central to social networks and sites for intimate relationships (Easthope, 2004). Households organize themselves around a variety of tasks, roles, and positions, creating a hierarchy that is significantly influenced by technology (Thoyre, 2020). With the advent of smart home technologies, new domains of domestic practices have emerged (Aagaard, 2023). For instance, Tolmie et al. (2007) explored the tasks and work involved in setting up and maintaining a networked home describing these activities as "digital housekeeping".

However, less attention has been paid to the ordinary and practical ways in which the evolving smart home devices are used on specific occasions. Instead, much of the available research has focused on what we might call 'broad brush' issues such as age stratification (e.g., Choi et al., 2019; Demiris et al., 2004; Zhang et al., 2009), disability (e.g., Jamwal et al., 2022; Mtshali & Khubisa, 2019), and so on. This has led, in our view, to a conspicuous gap in our understanding of the day-to-day dynamics of household life in these "smart" environments, and the various roles that the devices play in everyday practices, given the rapidly evolving nature of these technologies and new possibilities for their integrated use.

This gap is further highlighted as we observe the significant transformation in the social construction of living arrangements. The concept of home is evolving beyond a static physical space embracing fluid interaction with programmable devices that are becoming an integral part of our daily lives. In parallel, the construct of living arrangements is expanding to include various forms of co-living, such as co-housing and communal living, reflecting broader social and economic shifts. This evolving context highlights the critical need to examine boundary fluidity: the increasingly blurred lines between private and communal spaces, as well as between digital and physical realms. Understanding this fluidity is important for designing smart homes that not only adapt to the complexities of modern life and the diverse relationships of its inhabitants but also promote new forms of collaboration and interaction within these dynamic living environments.

Given the emergence of transformative agentic technologies—large language models, virtual reality, and dynamic avatars—it becomes urgent to discuss their integration into our homes and the opportunities they offer for navigating the complexities of domestic life. This workshop aims to foster collaboration between multidisciplinary researchers and practitioners, encouraging a joint effort to develop a taxonomy of the challenges and opportunities presented by these new, emerging technologies. By doing so, we strive to uncover insights that will guide the thoughtful design and implementation of technology, ensuring it harmonizes with and enriches our modern domestic lives.

## The Smart Home Ecosystem

There is little agreement on the definition of a smart home ecosystem; in fact, definitions vary according to the analytical lenses applied. Drawing from Gann et al.'s (1999) distinction between homes that simply contain smart appliances and those that allow interactive computing both within and beyond the home, Randall (2003) categorized smart homes into five types: (a) homes with intelligent stand-alone appliances, (b) homes where appliances exchange information to enhance functionality, (c) connected homes with internal and external networks for interactive control and access, (d) learning homes that record usage patterns to anticipate user needs, and (e) alert homes that monitor activities to proactively meet user needs . Taylor et al. (2007) highlighted that the 'smartness' of a home is not inherent in the devices themselves but emerges from how users integrate technology into their daily routines and activities. Advancing the discussion, Mennicken et al. (2015) proposed viewing smart homes as dynamic entities capable of evolving with users throughout their lives. Similarly, Reddy (2020) described the smart home as a Do-It-Yourself (DIY) 'process of be(com)ing with things,' during which the households appropriate, personalize, and customize their devices.

Building on this, our workshop focuses on systems that offer interactive elements, allowing users to actively construct their living spaces. This approach enables us to explore the intersection between people's routines and their practical use of new technologies. We aim to reveal how these interactions give birth to, or critically shape, the character and spirit of modern home living.

## Designing the Smart Home

"Home is a feeling", but a "smart home" as a notion has been contentious with people wanting technology to be more in the background since the advent of smart technologies for home environments (Eggen et al., 2003). This ongoing dialogue on how technology can augment home life's tasks, routines, and experiences has spurred extensive research (e.g., Aagaard, 2023; Mennicken et al., 2016; Woźniak et al., 2023). For example, Taylor et al. (2007) explored asynchronous communication within households and suggested the use of interactive artifacts, which was further expanded to enhance the connection among distant family members. Similarly, Jakobi et al. (2017) examined issues related to technology adoption, including the development of skills and the emergence of new technology-focused household responsibilities.

However, the HCI community's conceptual understanding of 'home'—its essence, location, creation, and creators—often seems limited. Typically, 'home' is viewed as a conventional house, while 'domestic life' is seen through the lens of family dynamics (Oogjes et al., 2018). A few pioneering studies have expanded the HCI discourse on home life by exploring the dynamics in unique settings such as subsistence communities, off-grid living, and cohousing communities (e.g., Jenkins, 2017; Leshed et al., 2014; Woodruff et al., 2008). Such research is critical to broadening perceptions of the notion of 'home' and fostering a richer understanding of the diversity of domestic life. Despite the value of these contributions, they remain few and far between.

The European Conference on Computer-Supported Cooperative Work provides an ideal setting for our workshop, aligning with the conference's focus on practice-oriented computing and the design of cooperative technologies. Our workshop aims to create a practice-oriented and reflexive environment to explore questions such as: 1) In what ways might new forms of agentic technologies, such as ChatGPT, robots, or virtual avatars, transform domestic life and what new practices might emerge from these changes? 2) What challenges and opportunities do these technologies introduce for collaborative and communal living spaces, such as gardens, meeting rooms, and hallways? 3) Specifically, how can we systematically categorize these challenges and opportunities into a taxonomy to better understand the integration of agentic technologies in home environments?

# Workshop Format

We are organizing a half-day event designed to engage participants in a meaningful exchange of thoughts and ideas regarding the challenges and opportunities presented by new technologies and their integration into the routines and practices of modern domestic life. We aim to host between 15 to 20 participants at most, with a minimum of 8, not including the organizers.

After setting up a workshop website, we will recruit via email lists (such as CSCW, CHI, Digital Culture, AOIR) and social media platforms (Facebook groups: SigCHI, Researchers of the sociotechnical, etc.; Twitter; Discord Channels; Slack Channels). Further, we will reach out within our respective networks, inside and outside of academia.

The important dates are:

Submission Deadline: April 22, 2024
Notification of Acceptance: May 10, 2024

Camera Ready Version Due: May 31, 2024

All deadlines are 23:59 anywhere on earth (AoE).

# Workshop Plan

We are planning an interactive workshop in which the participants will primarily engage in guided tasks to create a shared understanding of the current role technologies play in the everyday practices of domestic life. Our goal is to have pre- and post-workshop activities that will be tailored based on the submissions received and participants' desires and objectives for the workshop. We have created a tentative plan and schedule for the workshop day, which will be adjusted depending on the submissions received.

**Pre-Workshop Activities:**

- Posting questions for the group to reflect upon.
- Uploading submissions to Miro, look over the others and give feedback.
- Tentative (depending on group size): Informal online meetings to get to know one another.

**Preliminary Workshop Schedule:**

09:00–09:20: Welcome, Agenda, Intros.
09:20–9:50: Discussions in smaller Groups (Exploring how new forms of agentic technologies might transform domestic life and what new practices could emerge).
9:50–10:50: Discussions in smaller Groups (Exploring the challenges and opportunities of emerging technologies for modern domestic life).
10:50–11:15: Activity Break.
11:15–11:45: Sharing results of group work.
11:45–12:30: developing a taxonomy by categorizing the challenges and opportunities.
12:30–13:00: Next steps and closing.

Note: No special equipment is needed. We will likely need a moderation kit for the on-site participants to write, scribble, and draw on.

**Post-Workshop Activities:**

- Adding reflections from the workshop; engaging with others.
- Follow up announcement for the follow-up workshop.

## Planned Outcome

All the notes, documentation, and other materials that are created during the discussions will be shared among the workshop participants. We plan to organize follow-up workshops on other conferences to help this newly formed collaboration to continue, through discussions and new initiatives, thereby encouraging more researchers to reflect upon their own challenges when conducting research in home environment. We also plan to use the generated taxonomy of challenges and opportunities as a foundation for future work.

We aim at creating a network of researchers, practitioners, and designers engaging with the topic of future homes. We will set up a slack channel (for the workshop) that shall be used afterwards as well for sharing resources and planning collaborations.

## Call for Participation

We welcome and appreciate submissions in various formats, including traditional workshop papers, short essays, reflections (up to 4 pages, excl. references), video and audio recordings (max. 5 minutes), which focus on:

- The role of ChatGPT and similar models in creating new interactive possibilities
- The design of mobile or robotic installations which facilitate new interactional relationships
- The role of avatars
- Technologies at the boundary of the home, e.g., public spaces in co-living environments, such as gardens, meeting rooms, hallways, and so on.
- The impact of different lifestyle arrangements on smart home technology use
- Environmental considerations

Although this list is in no way intended to be exhaustive, creative and inspirational submissions would be particularly welcome. All submissions should come with a short bio of the applicant(s).

Submissions should be sent to Fatemeh.alizadeh@uni-siegen.de and will be reviewed based on relevance and potential for contribution to the workshop. At least one co-author of each accepted paper must register to the ECSCW 2024 conference to attend the workshop.

# Organizers

Fatemeh (Mahla) Alizadeh (main contact) and Dominik Pins are doctoral researchers in the field of HCI at the University of Siegen, Germany. Their research primarily explores users' sense-making of AI-based technologies and how to empower users in the face of technological limitations. Some of their publications include: "*I Don't Know, Is AI Also Used in Airbags? An Empirical Study of Folk Concepts and People's Expectations of Current and Future Artificial Intelligence*," published in the I-com Journal in 2021, "*Does Anyone Dream of Invisible AI? A Critique of the Making Invisible of AI Policing*" published in Nordic Human-Computer Interaction Conference in 2022, and "*Alexa, We Need to Talk: A Data Literacy Approach on Voice Assistants*," presented at the Designing Interactive Systems Conference 2021. They have also successfully co-organized a workshop at the ECSCW conference in 2022 on "*Building Appropriate Trust in Human-AI Interactions*" (Alizadeh, Vereschak, et al., 2022). In their recent project, SAM Smart (https://samsmart.de/), they explore the integration of smart home technologies into modern domestic life, aiming to design a home assistant that supports users in error handling and making sense of the collected data.

Dennis Paul is a research associate at Fraunhofer Institute for Applied Information Technology (FIT) in the department of Human-Centered Engineering and Design. In his quantitative UX research, he focuses on psychological factors that influence technology acceptance and user well-being.

Yuhui Xu is a Doctoral researcher in the field of HCI in the department of Industrial Design at Eindhoven University of Technology. His work focuses on care through home things. In his recent qualitative research, he explored designing a chatbot as an agent of everyday objects for mediating expats' loneliness in home contexts.

Mikołaj P. Woźniak is a doctoral researcher in human-computer interaction at University of Oldenburg, Germany. His work focuses on understanding smart homes as multi-user environments, with focus on empowering inhabitants in diagnosing and troubleshooting glitches with their domestic technology. In his recent work, he applies various qualitative methods to understand user strategies to cope with smart home malfunctions. Having his background in electronic engineering, Mikołaj also explored designing non-visual interfaces for smart home control. He has been a part of organizing committees for ISS, TEI, CHIWORK and CHI, responsible for volunteer coordination and hybrid technologies.

Sarah Mennicken is currently the founder of DREI Solutions, a UX consulting firm for startups focusing on user research, design and prototyping. She held industry research roles at Spotify and Microsoft Research where she focused on the integration of emergent technologies like conversational assistants, AR/VR, and interactive machine learning into everyday experiences. During her PhD, she explored user-centric smart home experiences and hosted workshops on the topic at CHI and Ubicomp.

Minha Lee is currently an assistant professor at the department of Industrial Design's Future Everyday group at Eindhoven University of Technology since 2020. Lee's expertise is in ethics and conversational user interfaces. She was the General co-chair of the ACM CUI Conference in Eindhoven in 2023, after being a Full papers and Provocation papers co-chair (2021, 2020), and was recently elected to chair the CUI Steering Committee. She has hosted workshops on CUIs and ethics at conferences like CHI, HRI, CSCW, and IUI.

Dave Randall is a senior professor at the University of Siegen. He has co-authored and edited eight books and in the order of 200 peer revied papers on a range of themes, in the main orienting to the role of qualitative research in various domains.  This includes one of the earliest studies of people actually living in a Smart home in 2003.

Peter Tolmie is Principal Research Scientist in the Information Systems and New Media group at the University of Siegen. He has co-authored and edited six books, with another two in preparation, and over 180 research papers and book chapters, tackling a diverse set of themes across the fields of HCI, CSCW, and Sociology. He is also Field Chief Editor for the Frontiers journal *Human Dynamics*. A great deal of his research from 2000 to 2015 was conducted in domestic environments and this forms the background to two of his books and over 25 of his published works.